\documentclass[runningheads]{llncs}
\usepackage[T1]{fontenc}
\usepackage{graphicx}
\usepackage{color}
\usepackage{amsmath,amssymb} 
\usepackage{hyperref}
\usepackage{algorithm}
\usepackage{algorithmic}
\usepackage{caption}
\usepackage{subcaption}
\usepackage{tikz}
\usepackage{float}

\usepackage{adjustbox}

\DeclareMathOperator*{\argmin}{arg\,min}
\DeclareMathAlphabet{\pazocal}{OMS}{zplm}{m}{n}
\newcommand{\unif}{\pazocal{U}}

\usepackage{mathtools}
\mathtoolsset{showonlyrefs=true}

\newcommand{\rev}[1]{\textcolor{black}{#1}}

\usetikzlibrary{arrows,shapes,shadows.blur,spy,fadings,shadings}
\tikzset{verre/.style={draw=SkyBlue,fill=SkyBlue!30}}
\tikzset{punkt/.style={rectangle,rounded corners,draw=black, very thick,text width=2cm,minimum height=2em,text centered}}
\tikzset{punkt2/.style={rectangle,rounded corners,draw=mygreen, very thick,text width=1.5cm,minimum height=2em,text centered}}
\tikzset{double arrow/.style args={#1 colored by #2 and #3}{
         -stealth,line width=#1,#2, 
          postaction={draw,-stealth,#3,line width=(#1)/3,
          shorten <=(#1)/3,shorten >=2*(#1)/3},}} 

\begin{document}
\title{Fluctuation-based deconvolution in fluorescence microscopy using plug-and-play denoisers}
\titlerunning{Fluctuation-based deconvolution using a PnP denoiser}

%

\author{Vasiliki Stergiopoulou\inst{1}\and
Subhadip Mukherjee\inst{2} \and
Luca Calatroni\inst{1}\and
Laure Blanc-Féraud\inst{1}
}
\authorrunning{V. Stergiopoulou et al.}

%
\institute{No institute given}
\institute{UCA, CNRS, INRIA, Laboratoire I3S, 06903, Sophia-Antipolis France \\
\email{vasiliki.stergiopoulou@i3s.unice.fr}, \email{calatroni@i3s.unice.fr}, \email{blancf@i3s.unice.fr} \and
Department of Computer Science, University of Bath, UK \\
\email{sm2467@cam.ac.uk}
}
\maketitle              
\begin{abstract}
\vspace{-0.7cm}
The spatial resolution of images of living samples obtained by fluorescence microscopes is physically
limited due to the diffraction of visible light, which makes the study  of entities of size less than
the diffraction barrier (around $200$ nm in the $x$-$y$  plane) very challenging. To overcome this limitation, several deconvolution and
super-resolution techniques have been proposed. Within the framework of inverse problems, modern approaches in fluorescence microscopy reconstruct a super-resolved image from a temporal stack of frames by carefully designing suitable hand-crafted sparsity-promoting regularisers. 
Numerically, such approaches are solved by  proximal
gradient-based iterative schemes. Aiming at obtaining a reconstruction more adapted to sample geometries (e.g.~thin filaments), we adopt a plug-and-play denoising approach with convergence guarantees and replace the proximity operator associated with the explicit image regulariser with an image denoiser (i.e. a pre-trained network) which, upon appropriate training, mimics the action of an implicit prior. To account for the independence of the fluctuations between molecules, the model relies on second-order statistics. The denoiser is then
trained on covariance images coming from data representing sequences of fluctuating fluorescent
molecules with filament structure. The method is evaluated on both simulated and real fluorescence
microscopy images, showing its ability to correctly reconstruct filament structures with  high values of peak signal-to-noise ratio (PSNR). 

\keywords{Fluorescence microscopy\and Image deconvolution \and Variational regularisation \and Proximal algorithms \and  Plug-and-Play regularisation.}
\end{abstract}
\section{Introduction}\label{sec: Introduction}


In optical microscopy, the highest achievable spatial resolution is governed by some fundamental physical laws related to light propagation and is therefore limited. According to Rayleigh's criterion, the resolution of an optical microscope is defined as the smallest resolvable distance, i.e.~the smallest distance between two point sources so that they can be distinguished in the image. For conventional fluorescence microscopes, this distance is approximately equal to $200$ nm in the lateral ($x$-$y$) plane.   In order to resolve sub-cellular structures of size smaller than this barrier, several deconvolution and super-resolution techniques have emerged in the literature. Originally developed in the applied fields of chemistry, biology, and biophysics, such techniques can be naturally described in more mathematical terms as regularisation approaches for solving the ill-posed inverse problem considered. A big family of approaches achieving nanometric resolution (around $20$ nm of lateral resolution) is known as \textit{Single Molecule Localisation Microscopy} (SMLM) techniques (see  \cite{SR_fight_club} for a review). These methods rely on the use of sparse regularisation approaches for reconstructing  frames of a temporal sequence of acquisitions where only a few molecules are active at a time. A more hardware-based super-resolution technique achieving a resolution of approximately $60-100$ nm, is the \textit{Stimulated Emission Depletion} (STED) microscopy approach \cite{Hell94} where the optical blur function (the microscope Point Spread Function, PSF) is depleted by means of suitable devices. While effective, these techniques show several drawbacks: for example, SMLM has  long acquisition times while STED requires highly expensive commercial tools. Furthermore, both approaches require special (and expensive) fluorescent molecules able to support the high laser power required. 

For overcoming such limitations, different types of approaches exploiting the independent stochastic temporal fluctuations of distinct fluorescent emitters became popular over the last decade. 
Such approaches rely on the use of both standard microscopes and fluorescent molecules and represent therefore a powerful class of approaches for applications. Some of these approaches are: the \textit{Super-resolution Optical Fluctuation Imaging} (SOFI) approach \cite{sofi} where the lack of correlation between distinct emitters is exploited by analyzing high-order statistics, the \textit{Super-Resolution Radial Fluctuations} (SRRF) \cite{srrf} microscopy, where super-resolution is achieved by calculating at each frame the degree of local symmetry and, finally, the \textit{Sparsity-based Super-resolution Correlation Microscopy} (SPARCOM) \cite{SPARCOM} which models the sparse distribution of the fluorescent molecules via the use of a convex $\ell_1$ regularisation applied on the emitters' covariance matrix.
To improve the performance of SPARCOM, the Covariance-based $\ell_0$ super-resolution microscopy with intensity estimation (COL0RME) method \cite{stergiopoulou_BioIm,Stergiopoulou_ISBI21} has been proposed to estimate both molecule positions and intensities, which is a valuable piece of information in several applications, such as, e.g., 3D imaging \cite{Stergiopoulou_ISBI22}, by means of a two-step procedure relying on hand-crafted sparsity promoting
regularisers. The approach further estimates noise statistics and background terms (containing out-of-focus molecules). 
Both SPARCOM and COL0RME rely on the minimization of non-smooth (and possibly non-convex) functionals, for which tailored proximal optimization algorithms \cite{Proximal_Algorithms} based on soft- and hard-thresholding rules have thus been considered.

The applicability of  these approaches to more complex geometries is limited due to the hand-crafted sparsity they enforce which creates biases (i.e., punctuated structures) in the reconstruction. 
This is particularly limiting when continuous curvilinear structures are desired, which is the case in several biological applications.
For that, suitable regularisers can indeed be defined \cite{Laville2022}, with the major limitation of remaining  tailored to particular shapes only. With the intent of developing a flexible regularisation approach suited to adapt to different geometries, we present in the following a data-driven optimization-inspired technique relying on the use of the so-called Plug-and-Play (PnP) approaches \cite{PnP_2013}, which, over the last decade have been proved to represent an efficient framework for solving inverse image restoration problems, see \cite{kamilov2022plug} for a review. In this framework, the regulariser is parameterised by a deep neural network that can be trained on simulated data implicitly characterised by desired structures of interest, thus better capturing/promoting their shape after training. Our primary motivations behind using the PnP approach are three-fold: (i) Training the denoiser is independent of the imaging forward operator, which makes the pre-trained denoiser applicable even if the forward operator undergoes some changes, without having to retrain the denoiser from scratch. (ii) The training problem does not require pairs of measured and ground-truth images, unlike supervised machine learning approaches. To train the denoiser, one only needs high-quality ground-truth images and their noisy counterparts (with additive Gaussian noise).  (iii) The PnP approaches are rooted in proximal point algorithms, so their convergence can be rigorously studied using results from fixed-point theory and/or convex analysis. This leads to better interpretability of the reconstruction algorithm and results in a principled way of combining knowledge about imaging physics with the available training data.

\paragraph{Contributions:} In this work, we leverage the framework of PnP approaches with convergence guarantees \cite{cohen2021,hurault2022gradient,Hurault2022proximal} to show good empirical performance on  the inverse problem of fluctuation-based image deconvolution presented, e.g., in \cite{SPARCOM,stergiopoulou_BioIm,Stergiopoulou_ISBI21}. 
In Section \ref{sec:pnp}, we review the recent advances in the field of PnP approaches for inverse problems, pointing out the convergent scheme we employ. In Section \ref{sec:col0rme}, we formulate the covariance-based deconvolution model and formulate its PnP extension, which we called PnP-COL0RME in the following. In Section \ref{sec: Results PnP-COl0RME} we report several numerical results on both simulated and real data where the advantages of using the PnP reconstruction model are shown in comparison to its model-based counterpart.

\section{Plug-and-Play approaches for inverse problems}  \label{sec:pnp}

A standard approach for solving ill-posed inverse problems in imaging consists in solving the optimisation problem:
\begin{equation}\label{eq:regularized inverse problem}   
\hat{\mathbf{x}} \in \argmin_{\mathbf{x}\in\mathbb{R}^{\rev{n^2}}} ~\mathcal{F}(\mathbf{\Psi x};\mathbf{y}) +\lambda \mathcal{R}(\mathbf{x}), \quad \lambda>0,
\end{equation}
where, for observed data $\mathbf{y}\in\mathbb{R}^{n^2}$ (being the vectorisation of a 2D image of size $n\times n$) and model operator $\mathbf{\Psi}\in\mathbb{R}^{n^2\times n^2}$, $\mathcal{F}$ denotes a (smooth) data fidelity term and $\mathcal{R}$ a regularisation term encoding prior knowledge on the solution $\hat{\mathbf{x}} \in\mathbb{R}^{n^2}$.
Depending on the available prior information (such as sparsity, gradient smoothness, etc.), tailored hand-crafted functions $\mathcal{R}$ can be used. 
In most cases, $\mathcal{R}$ is non-smooth, and proximal algorithms \cite{Proximal_Algorithms} can be used for solving \eqref{eq:regularized inverse problem}. We recall that the proximity operator of parameter $\tau>0$ of a proper, convex and non-smooth function $\mathcal{R}$ is defined by:
\begin{equation}\label{proxim_operator}
    \text{prox}_{\tau \mathcal{R}}(\mathbf{z}) = \argmin_{\mathbf{x}\in\mathbb{R}^{n^2}} ~\mathcal{R}(\mathbf{x}) + \frac{1}{2\tau} \|\mathbf{z}-\mathbf{x} \|_2^2,\quad \mathbf{z}\in\mathbb{R}^{n^2}.
\end{equation}
Solving \eqref{proxim_operator} corresponds to solving the problem of denoising an image $\mathbf{z}\in\mathbb{R}^{n^2}$ corrupted by an additive white Gaussian noise (AWGN) of constant variance equal to $\tau$. Within a proximal gradient algorithm, \eqref{proxim_operator} can thus be interpreted as a denoising step of the gradient descent iteration $\mathbf{z}\rev{^k}=\mathbf{x}^k-\tau\nabla \mathcal{F}(\mathbf{\Psi x^k};\mathbf{y})$ at each iteration $k\geq 0$. This observation inspired the authors of \cite{PnP_2013} to develop the framework of PnP priors, whose main idea \rev{consists in replacing} $\text{prox}_{\tau \mathcal{R}}(\cdot)$ with an off-the-shelf image denoiser $D_\sigma(\cdot)$ depending on a parameter $\sigma>0$ corresponding to a regularisation functional $\mathcal{R}$ whose explicit definition is often not available. In a Bayesian framework, it is indeed possible to  explicitly relate Gaussian minimum mean-squared error (MMSE) denoisers $D_\sigma(\cdot)$  with the (unknown) image prior $p(\cdot)$ one would like to model \cite{Laumont2022} using the Tweedie's identity: $\sigma^2 \nabla  \log p_\sigma(\mathbf{x}) = D_\sigma(\mathbf{x})-\mathbf{x}$,
where $p_\sigma$ is the convolution of $p$ with a Gaussian smoothing kernel of bandwith $\sigma>0$, which makes $p_\sigma$ smoother (namely, Lipschitz differentiable) than $p$ under mild conditions. \rev{As observed in \cite[Eq. 74-75]{Reehorst2019}, considering a (Gaussian) denoiser residual is in fact a good approximation to the score of the image prior.} Along with their Bayesian interpretability, another advantage of PnP approaches is that they allow the use of advanced image denoising models, e.g. denoisers parameterised by convolutional neural networks (CNNs), within the iterative scheme, with impressive representational capabilities. In most cases, the CNN image denoiser $D$ is trained to perform denoising on some pairs of clean-noisy images and can be used afterward for more-general inverse problems (e.g. deblurring, super-resolution, etc.), see \cite{kamilov2022plug} for a review. Some state-of-the-art denoisers include image-dependent filtering algorithms such as Block-Matching \& 3D filtering (BM3D) \cite{BM3D},  Denoising Convolutional Neural Networks (DnCNN) \cite{zhang2017beyond} and Dilated-Residual U-Net (DRUNET) deep learning network \cite{zhang2021plug}.

These denoisers are typically used in iterative proximal schemes (see, e.g., \cite{Kamilov2017} for a FISTA-type PnP scheme), although they can be flexibly used in other algorithms such as, e.g., the
Alternate Directions Method of Multipliers (ADMM) \cite{BoydADMM}, the Douglas-Rachford Splitting (DRS) \cite{Douglas_Rachford} and the Half-Quadratic Splitting (HQS) \cite{HQS}. For all these algorithms, a corresponding PnP version can indeed simply be obtained as described above. PnP versions of proximal algorithms have been used to solve image restoration problems such as for example PnP-PGD in \cite{terris2020}, PnP-ADMM and PnP-DRS in \cite{RED,ryu19a} and PnP-HQS in \cite{zhang2017beyond,zhang2021plug,cohen2021}. 


In \cite{RED} an explicit regularisation by denoising (RED) strategy was designed in terms of an explicit function $\mathcal{R}(\cdot)$ defined, for generic image denoiser $D$, by:
\begin{equation}
     \mathcal{R}(\mathbf{x}) := \frac12 \mathbf{x}^{T} (\mathbf{x} - D(\mathbf{x})).
\end{equation}
Under conditions of local homogeneity, non-expansiveness, and Jacobian symmetry, $D$ was shown to be indeed equivalent to a gradient step on $\mathcal{R}$ \cite{Reehorst2019}, that is, $D(\mathbf{x}) = \mathbf{x} - \nabla \mathcal{R}(\mathbf{x})$.
However, as shown in \cite{Reehorst2019}, such requirements are unrealistic on the widely-used denoisers mentioned above, as they do not have symmetric Jacobians. 
In order to overcome this limitation, in \cite{cohen2021,hurault2022gradient,Hurault2022proximal}, the authors proposed to formulate, similar to RED, a gradient step denoiser of the form:
\begin{equation}\label{denoiser hurault}
    D_\sigma(\mathbf{x}) = \mathbf{x} - \nabla \mathcal{R}_\sigma(\mathbf{x}),
\end{equation}
where $\mathcal{R}_\sigma:\mathbb{R}^{n^2}\rightarrow \mathbb{R}$ is a scalar function parameterised by a neural network $N_\sigma:\mathbb{R}^{n^2}\rightarrow \mathbb{R}^{n^2}$. Interestingly, under mild structural assumption on $D_\sigma$, the authors are able to prove sound convergence guarantees for the underlying non-convex optimisation problem defined in terms of a non-trivial (but explicit) regularisation function $\mathcal{R}(\cdot)$.  In the following section we specify the particular problem we are interested in and discuss its PnP extension based on the strategy discussed above.


\section{Deconvolution via sparse auto-covariance analysis}\label{sec:col0rme}

We consider the following image formation model considered, e.g., in \cite{SPARCOM,Stergiopoulou_ISBI21,stergiopoulou_BioIm} to describe, for $t=1,\ldots, T, ~T>0$, a video of temporal acquisitions $\mathbf{y}_t \in \mathbb{R}^{n^2}$ \rev{by standard fluorescent microscopes of true images $\mathbf{x}_t \in \mathbb{R}^{n^2}$}:
\begin{equation}\label{forward model simplified PnP}
    \mathbf{y}_t = \mathbf{\Psi} \mathbf{x}_t +
    \mathbf{b}+\mathbf{n}_t.
\end{equation}
In \eqref{forward model simplified PnP}, $\mathbf{\Psi}\in \mathbb{R}^{n^2\times n^2}$ is a (known) convolution operator associated with the system point spread function (PSF),  $\mathbf{b}\in \mathbb{R}^{n^2}$ is a background term and $\mathbf{n}_t \in \mathbb{R}^{n^2}$ is the realisation at time $t$ of an i.i.d.~Gausian noise random vector with unknown variance $s \geq 0$, i.e. 
 $\mathbf{n}\sim\mathcal{N}(0,s\,\mathbf{Id})$. \rev{We look for a deconvolved image $\mathbf{x}\in\mathbb{R}^{n^2}$, defined as $\mathbf{x}= \frac{1}{T}\sum_{t=1}^T \mathbf{x_t}$.} In \cite{SPARCOM,stergiopoulou_BioIm}, a reformulation of the model \eqref{forward model simplified PnP} was done in the covariance domain in order to exploit temporal information. 
In the following, we proceed similarly but consider a simplified modeling where only auto-covariance vectors are taken into account, thus neglecting cross-terms. 

 Considering the frames $\{\mathbf{y}_t\}_{t=1}^T$ as $T$ realisations of a random variable $\mathbf{y}$, the sample auto-covariance \rev{(variance)} vector $\mathbf{\tilde{r}_y}\in\mathbb{R}^{n^2}$ of $\mathbf{y}$ can be estimated by:
 \begin{equation}\label{eq:cov_mat_emperical_auto}
    \mathbf{\Tilde{r}_y} \approx
    \frac{1}{T-1}\sum_{t=1}^T (\mathbf{y}_t-\overline{\mathbf{y}})^2,
\end{equation}
where $\overline{\mathbf{y}}=\frac{1}{T}\sum_{t=1}^T \mathbf{y}_t$ denotes the empirical  mean. 
%
 %
%
%
%
From \eqref{forward model simplified PnP} and \eqref{eq:cov_mat_emperical_auto}, the following model thus holds between the auto-covariance vectors:
\begin{equation}\label{eq: model auto-covariance}
    \mathbf{\Tilde{r}_y} = \mathbf{\Psi}^2 \mathbf{r_x} + \mathbf{\Tilde{r}_n},
\end{equation}
where $\mathbf{r_x} \in \mathbb{R}^{n^2}$ and $\mathbf{\Tilde{r}_n} \in \mathbb{R}^{n^2}$ are the auto-covariance vectors associated to the samples $\{\mathbf{x}_t \}_{t=1}^T$ and $\{\mathbf{n}_t\}_{t=1}^T$, respectively, and where
by $\mathbf{\Psi}^2 \in \mathbb{R}^{n^2 \times n^2}$ we denote the matrix $\mathbf{\Psi}\odot \mathbf{\Psi}$ where $\odot$ denotes the point-wise Hadamard product. Finally, note that by assumption $\mathbf{\Tilde{r}_n} = s \mathbf{1}$, where $\mathbf{1} = (1,\ldots,1)\in\mathbb{R}^{n^2}$.

\begin{remark}
Note that, upon reshaping, $\mathbf{\Tilde{r}_y}\in\mathbb{R}^{n^2}$ is in fact the second-order SOFI image associated to the stack  $\{\mathbf{y}_t \}_{t=1}^T$, which, thanks to the `squaring' of the underlying point spread function, enjoys better spatial resolution in comparison, e.g., to $\bar{\mathbf{y}}$, see \cite{sofi} for details. 
\end{remark}

\begin{remark}
In comparison to the covariance-based modelling in SPARCOM \cite{SPARCOM} and COL0RME \cite{stergiopoulou_BioIm}, model \eqref{eq: model auto-covariance} is indeed a simplification. In those papers, the whole sample-covariance matrix $\mathbf{{R_y}}\in\mathbb{R}^{n^2\times n^2}$ with main diagonal $\mathbf{\Tilde{r}_y}$ was computed. Such a matrix is not diagonal, due to the correlation induced by the PSF. In order to deal with a simplified model and benefit from faster calculations, we consider in the following the relation \eqref{eq: model auto-covariance} involving only auto-covariance terms and leave a complete modelling involving also cross terms for future work. The resulting observation model \ref{eq: model auto-covariance} is thus less rich but exact (not approximated).
\end{remark}

Based on \eqref{eq: model auto-covariance}, we are now interested in finding the fluorescent molecule locations and estimate noise information. Namely, we are interested in finding the support of $\mathbf{r_x}$,  $\mathbf{\Omega}:=\left\{ i~:~(\mathbf{r_x})_i \neq 0 \right\}$ and the unknown noise variance $s\geq 0$.  We do so by considering the following minimisation problem:
\begin{equation} \label{eq: support estimation auto}
    (\hat{\mathbf{r}}_\mathbf{x}, \hat{s})\in\argmin\limits_{\mathbf{r_x} \geq 0,~ s \geq 0}~ \frac{\lambda}{2} \| \mathbf{\Tilde{r}_y} -\mathbf{\Psi^2} \mathbf{r_x} - s \mathbf{1} \|_2^2 +  \mathcal{R}(\mathbf{r_x}),
\end{equation}
where $\lambda>0$ is a regularization parameter and $\mathcal{R}(\cdot)$ is a regularisation term to be defined to enforce desirable properties (sparsity, for instance) of the solution. Problem \eqref{eq: support estimation auto} can be solved by Algorithm \ref{Algorithm:AMA_support_auto}, \rev{where, to improve convergence speed, a global minimisation on $s$ is performed followed by a proximal-gradient step on $\mathbf{r_x}$. An analogous (a priori, slower) algorithm benefiting from theoretical convergence guarantees is the Proximal Alternating Linearized Minimisation (PALM) algorithm whose convergence is studied in \cite{Bolte2014}.}

\begin{algorithm}[t]
\caption{ Model-based and PnP support estimation}
\label{Algorithm:AMA_support_auto}
\begin{algorithmic}
\REQUIRE $\mathbf{\Tilde{r}_y}, \mathbf{r_x}^0\in\mathbb{R}^{n^2}$ and parameters ($\tau,\lambda>0$ for model-based, $\sigma$ for PnP)
\REPEAT 
\vspace{0.2cm}
\STATE $s^{k+1} =  \mathbf{1}^T ( \mathbf{\Tilde{r}_y} -\mathbf{\Psi^2} \mathbf{r_x}^k)$
\vspace{0.2cm}
\STATE $\mathbf{z}^{k+1} = \mathbf{r_x}^{k} - \tau\lambda (\mathbf{\Psi^2})^T (\mathbf{\Tilde{r}_y} -\mathbf{\Psi^2} \mathbf{r_x}^k - s^{k+1} \mathbf{1}) $
\vspace{0.2cm}
\STATE  $\mathbf{r_x}^{k+1} = \begin{cases}
\text{prox}_{\tau\lambda R}(\mathbf{z}^{k+1})\quad &\quad \texttt{\% model-based}~\eqref{eq: support estimation auto}\\
\mathbf{r_x}^{k+1}=D_\sigma(\mathbf{z}^{k+1})\quad &\quad\texttt{\% PnP}
\end{cases}
$
\vspace{0.2cm}
\UNTIL convergence
\RETURN $\mathbf{\Omega}:=\left\{ i~:~(\mathbf{\hat{r}_x})_i \neq 0 \right\}, \hat{s}$
\end{algorithmic}
\end{algorithm}


Once $\mathbf{\Omega}$ and $\hat{s}$ have been computed, following \cite{Stergiopoulou_ISBI21,stergiopoulou_BioIm} a second algorithmic step can be performed to estimate image intensities only in correspondence with the support points in $\mathbf{\Omega}$, i.e. by solving:
\begin{equation} \label{eq: intensity estimation}
    (\hat{\mathbf{x}}, \hat{\mathbf{b}})\in\argmin\limits_{\mathbf{x}\in\mathbb{R}_+^{|\Omega|},~ \mathbf{b}\in\mathbb{R}_+^{n^2}} ~\frac12 \|\mathbf{\Psi_\Omega} \mathbf{x} - (\overline{\mathbf{y}} - \mathbf{b})\|_2^2 + \frac{\mu}{2} \|\nabla_\Omega\mathbf{x}\|_2^2 + \frac{\beta}{2} \|\nabla\mathbf{b}\|_2^2,
\end{equation}
where the data term models the presence of Gaussian noise, 
$\overline{\mathbf{y}} = \sum_{t=1}^T\mathbf{y_t}$ and $\mu,\beta>0$ are regularisation parameters. \rev{Moreover, $\mathbf{\Psi_\Omega} \in\mathbb{R}^{n^2\times |\Omega|}$ is a matrix whose $i$-th column is extracted from $\mathbf\Psi$ for all indexes $i \in \Omega$ and $\nabla_\Omega$ denotes the discrete gradient operator restricted to points in the support $\Omega$.}


 A hand-crafted regularisation model $\mathcal{R}$ in \eqref{eq: support estimation auto} introduces  reconstruction biases. For example, using the $\ell_1$ norm \cite{SPARCOM} or the continuous exact relaxation of the $\ell_0$ pseudo-norm \cite{Soubies-et-al-2015}) enforces sparsity by promoting point reconstruction. Solutions thus appear dotted as reconstructed points have a given inter-distance which cannot be decreased \cite{Duval-Peyre2015}. For reconstructing filaments, a solution is to use a regularising term promoting curves. Such method is proposed, e.g., in \cite{Laville2022} in an off-the-grid setting, but the numerical aspects are difficult and still under development. To overcome this limitation, in the following section, we propose a Plug-and-Play extension of the approach above where the proximal step is replaced by a denoiser $D_\sigma$ trained on an appropriate dataset of covariance images representing the geometrical structures of interest.


\subsection{Plug-and-play extension}\label{sec:pnp_col0rme}

The proximal step naturally appearing when solving problem \eqref{eq: support estimation auto} by proximal gradient algorithms, can be replaced by an off-the-shelf denoiser. To do so, we make use of a proximal gradient step denoiser as proposed by Hurault \textit{et al.} in \cite{Hurault2022proximal}. In their paper, the authors showed that this choice corresponds indeed to the proximal operator associated to a non-convex smooth function which allows the authors to derive convergence guarantees of the resulting proximal gradient scheme \cite[Theorem 4.1]{Hurault2022proximal}. 
 Note, that differently to the setting proposed in \cite{Hurault2022proximal}, our algorithm processes auto-covariance images due to the model \eqref{eq: model auto-covariance} and, along with $\hat{\mathbf{r}}_\mathbf{x}$, it provides an estimate $\hat{s}$ of the noise variance by alternate minimisation. We report the iterative scheme in Algorithm \ref{Algorithm:AMA_support_auto} and refer in the following to PnP-COL0RME to the case when a PnP regulariser is employed.
 


In \cite{Hurault2022proximal} the authors considered a denoiser $D_\sigma$  in the form of a gradient step \eqref{denoiser hurault} of a functional \rev{$\mathcal{R}_\sigma : \mathbb{R}^{n^2} \rightarrow \mathbb{R}$ with specific properties, e.g., bounded from below, and} parameterised by a deep neural network $N_\sigma$. Recalling the characterisation of proximity operators \cite{Gribonval2020} introduced by Gribonval \& Nikolova, the authors proved in fact that $D_\sigma$ can be written as proximal operator of a function $\phi_\sigma$ defined by:
\begin{equation}\label{function phi}
    \phi_\sigma(\mathbf{w}) := \mathcal{R}_\sigma(D_\sigma^{-1}(\mathbf{w}))-\frac{1}{2}\|D_\sigma^{-1}(\mathbf{w})-\mathbf{w}\|_2^2,\quad\mathbf{w}\in\mathbb{R}^{n^2}.
\end{equation}
The function minimised when employing PnP COL0RME reads: $F_{\sigma}(\mathbf{r_x},s):=  \frac{1}{2} \| \mathbf{r_y} -\mathbf{\Psi^2} \mathbf{r_x} - s \mathbf{1}\|_2^2+\phi_\sigma(\mathbf{r_x})$,
which, after recalling that  $\mathbf{r}^k = D_\sigma(\mathbf{z}^k)$ at each $k$, can be written as:
\begin{equation} \label{function PnP2}
    F_{\sigma}(\mathbf{r_x}^k,s^k)= \frac{1}{2} \| \mathbf{r_y} -\mathbf{\Psi^2} \mathbf{r_x}^k - s^k \mathbf{1}\|_2^2+ \mathcal{R}_\sigma(\mathbf{z}^k)-\frac12\|\mathbf{z}^k-\mathbf{r_x}^k\|_2^2.
\end{equation}
In \cite[Theorem 4.1]{Hurault2022proximal} the authors show that thanks to the structure of $F_\sigma$, the PnP proximal gradient scheme converges indeed to a stationary point of $ F_{\sigma}$, whose decay can be indeed assessed throughout the iterations.

Note that, the regularisation parameter $\lambda>0$ appearing in \eqref{eq:regularized inverse problem} to regulate the strength of the regularisation term $\mathcal{R}$ has  been replaced by the hyperparameter $\sigma$ in \eqref{function PnP2}.  Intuitively, the value of $\sigma$ should correspond to the variance of AWGN appearing in the gradient steps of the proximal gradient algorithm \ref{Algorithm:AMA_support_auto}, hence its tuning is not straightforward.
As discussed in \cite{scaling_parameter}, a possible remedy for avoiding a time-consuming parameter tuning consists in introducing a rescaling parameter whose setting is easier than $\sigma$. 

\section{Numerical results}\label{sec: Results PnP-COl0RME}

We now present some results obtained by using PnP-COL0RME on temporal sequences of blurred and noisy data. A natural extension to the actual problem of super-resolution where $\mathbf{\Psi}=\mathbf{M_q}\mathbf{H}\in\mathbb{R}^{m^2\times n^2}$ with $\mathbf{H}\in\mathbb{R}^{n^2\times n^2}$ is PSF convolution matrix and $\mathbf{M_q}\in\mathbb{R}^{m^2\times n^2}$ is a downsampling operator with $n=qm, q>1$, is left for future work.

\begin{figure}[h!]

     \centering
     \begin{subfigure}[b]{0.3\textwidth}
         \centering
         \includegraphics[width=\textwidth]{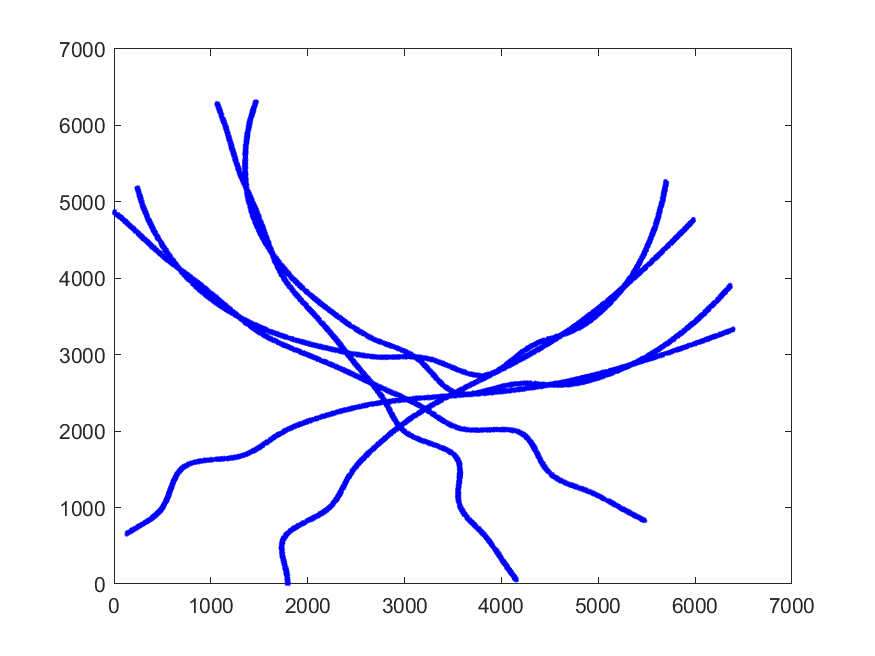}
     \end{subfigure}
     \hfill
      \begin{subfigure}[b]{0.3\textwidth}
         \centering
         \includegraphics[width=\textwidth]{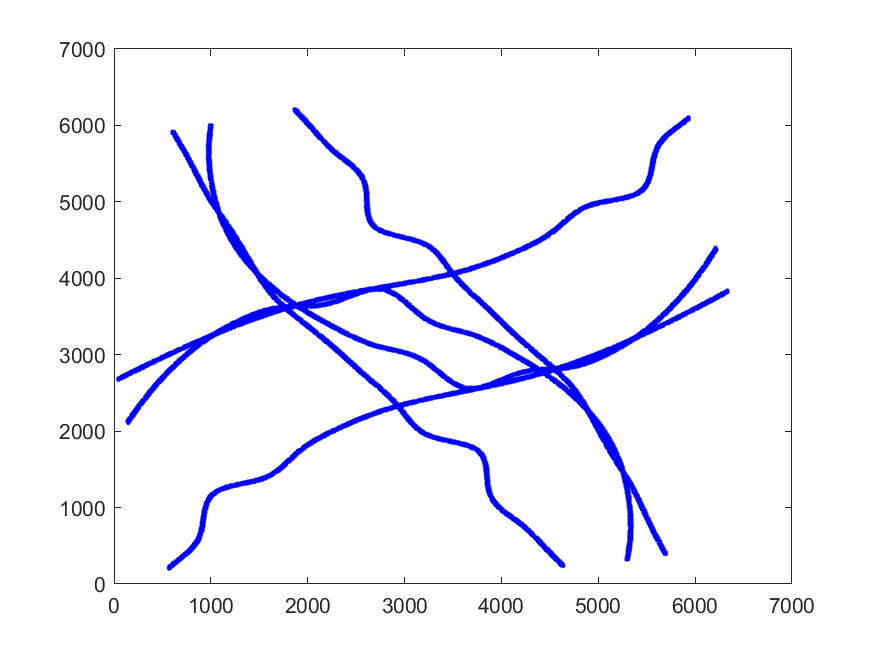}
     \end{subfigure}
\hfill    
      \begin{subfigure}[b]{0.3\textwidth}
         \centering
         \includegraphics[width=\textwidth]{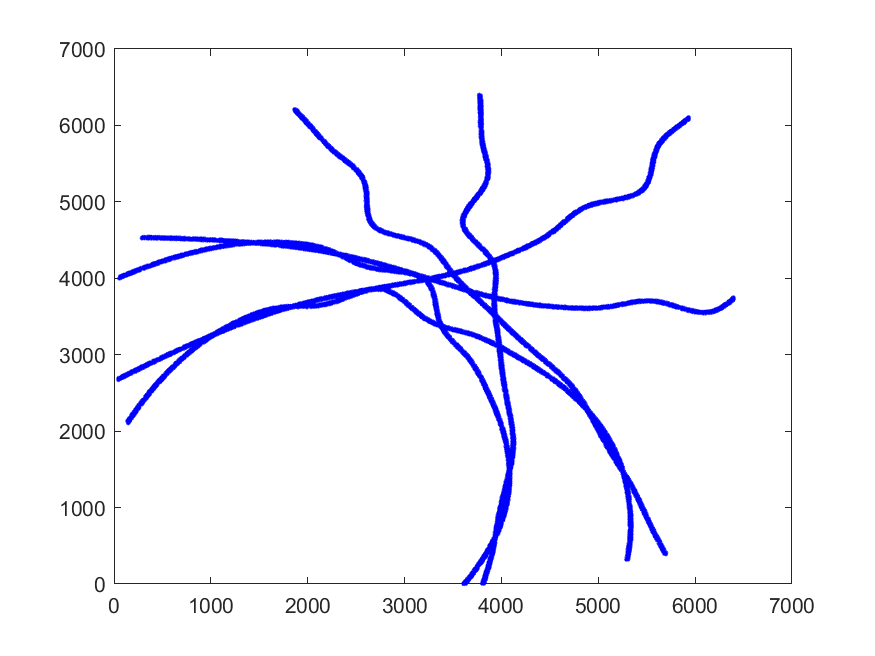}
     \end{subfigure}
     \caption{Simulated spatial patterns}
     \label{fig: dif spatial}
 \end{figure}
\begin{figure}[h!]
       \begin{subfigure}[b]{0.3\textwidth}
         \centering
         \includegraphics[height=3cm]{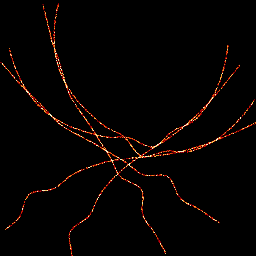}
         \caption{}
         \label{a:fluct}
     \end{subfigure}
     \hfill
      \begin{subfigure}[b]{0.3\textwidth}
         \centering
         \includegraphics[height=3cm]{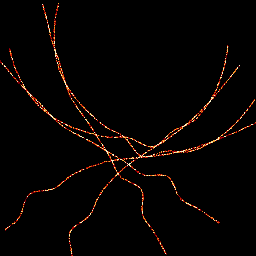}
         \caption{}
         \label{b:fluct}
     \end{subfigure}
      \hfill
      \begin{subfigure}[b]{0.3\textwidth}
         \centering
         \includegraphics[height=3cm]{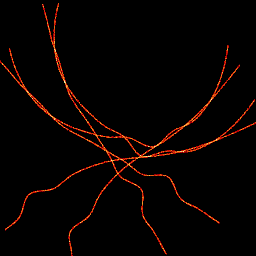}
       \caption{}
       \label{c: fluct}
     \end{subfigure}
     \caption{(a-b) Two different frames of a simulated fluctuating stack made from the first spatial pattern from Figure \ref{fig: dif spatial}, (c) The auto-covariance image $\mathbf{r_x^{\text{GT}}}$ estimated from the whole temporal sequence.}
\end{figure}

To train the denoiser $D_\sigma$ we created a dataset composed of clean and noisy image pairs. The geometrical features of the images in this dataset should be the same as the one of the images to restore. Differently from other methods, the proposed algorithm works with a model formulated in the covariance domain, so that the denoiser takes as an input noisy sample auto-covariance matrices of a fluctuating temporal sequence of images. Hence,  to create the dataset we first started by creating different spatial patterns (thin filaments) shown in Figure \ref{fig: dif spatial} where the emitters have different positions in the continuous grid. Such patterns are the superposition, after rotations with different angles, of the ground truth spatial pattern provided in the MT0 microtubule training dataset uploaded for SMLM 2016 challenge \footnote{\href{https://srm.epfl.ch/Challenge/ChallengeSimulatedData}{https://srm.epfl.ch/Challenge/ChallengeSimulatedData}}. Then, we used the fluctuation model discussed in \cite{SOFItool} to simulate temporal fluctuations and create a temporal stack of $T=500$ frames for each spatial pattern. Two exemplar frames of one temporal stack of images are reported in Figures \ref{a:fluct} and \ref{b:fluct}. For each temporal stack of images, we could therefore calculate the temporal auto-covariance image (see Figure \ref{c: fluct}) corresponding to one instance of the clean images $\mathbf{r_x^{\text{GT}}}$ in our dataset. To create now its noisy version we added Gaussian noise $\boldsymbol\eta$ with constant variance $\sigma^2$, $\boldsymbol\eta\sim\mathcal{N}(\mathbf{0},{\sigma^2}\mathbf{Id})$, with $\sigma$ following a uniform distribution, $\sigma \sim \unif(\sigma_1,\sigma_2)$.
We remark that since the noise in the covariance data comes from additive Gaussian noise on the individual frames, its actual distribution is indeed $\chi^2$. However, since the number of the degrees of freedom is high (as $T=500$), the distribution can be  approximated by a Gaussian distribution.
In our experiments, after normalising $\mathbf{r_x^{\text{GT}}}$ with maximum value equal to $1$, we select $\sigma_1= \epsilon<<1$ and $\sigma_2 = 50/255$.

Training was performed following the procedure in \cite{Hurault2022proximal} and using the code available on the authors' GitHub repository \footnote{\href{https://github.com/samuro95/Prox-PnP}{https://github.com/samuro95/Prox-PnP}}. For the neural network $N_\sigma(\cdot)$ used to parameterise the denoiser (see \eqref{denoiser hurault}), we used DRUNet, a CNN proposed in \cite{zhang2021plug}. For training, we used $500$ pairs of clean-noise auto-covariance images and $100$ for validation. The network was trained using $1215$ epochs via ADAM optimization and batch size equal to $16$. In the following experiments, the choice $\tau=1$ and $\lambda=0.99$ in Algorithm \ref{Algorithm:AMA_support_auto} was performed to guarantee convergence, see \cite[Section 4.1]{Hurault2022proximal} for details.

\subsection{Simulated data}\label{PnP simulated data}

\begin{figure}[h!]

 \begin{subfigure}[b]{0.24\textwidth}
     \centering
     \includegraphics[height=2.8cm,trim={5cm 1cm  3cm 1cm},clip]{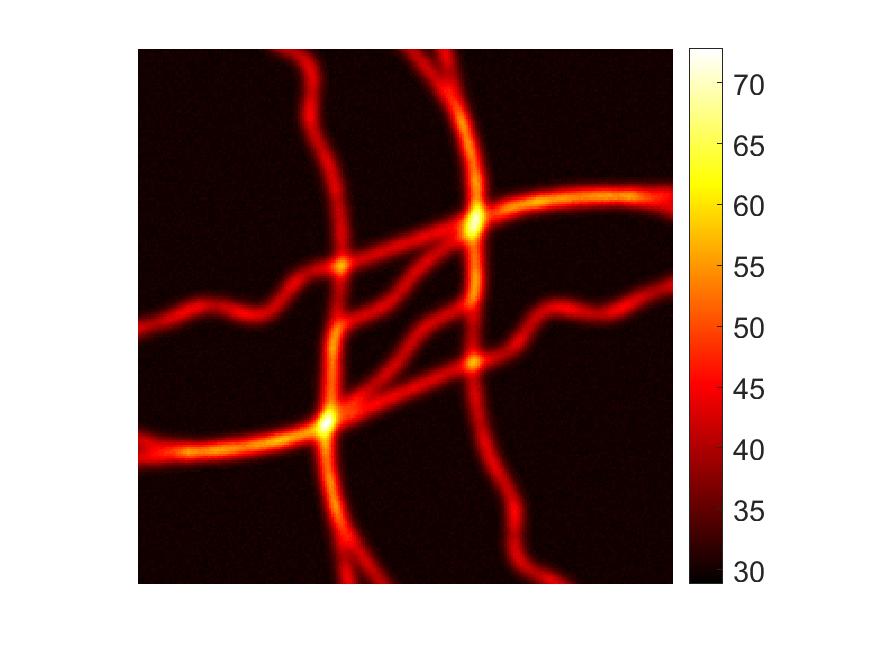}
     \caption{$\overline{\mathbf{y}}$}
 \end{subfigure}
 \hfill
 \begin{subfigure}[b]{0.24\textwidth}
     \centering
     \includegraphics[height=2.8cm,trim={5cm 1cm  3cm 1cm},clip]{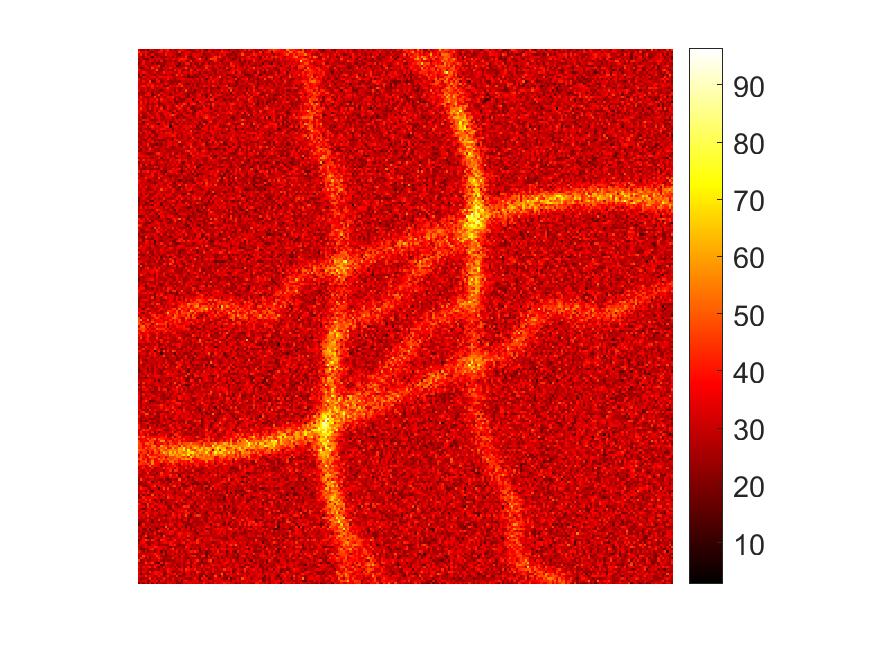}
     \caption{$\mathbf{y_1}$}
     \label{fig: bg_blur}
 \end{subfigure}
\hfill
 \begin{subfigure}[b]{0.24\textwidth}
     \centering
     \includegraphics[height=2.8cm,trim={5cm 1cm  7cm 1cm},clip]{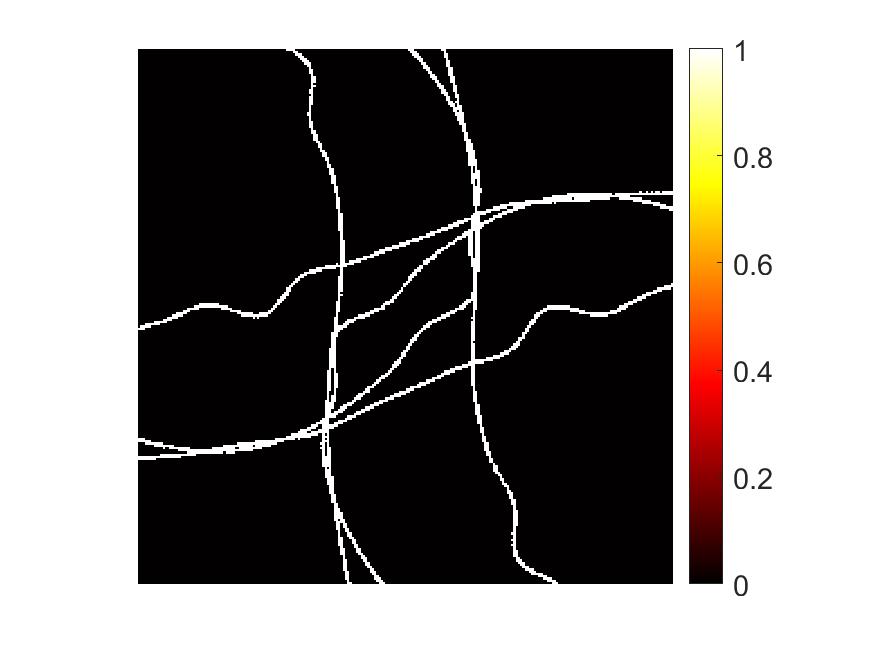}
     \caption{$\mathbf{\Omega_{\text{GT}}}$}
     \label{fig:supp_GT}
 \end{subfigure}
 \hfill
 \begin{subfigure}[b]{0.24\textwidth}
         \centering
         \includegraphics[height=2.8cm,trim={5cm 1cm  3cm 1cm},clip]{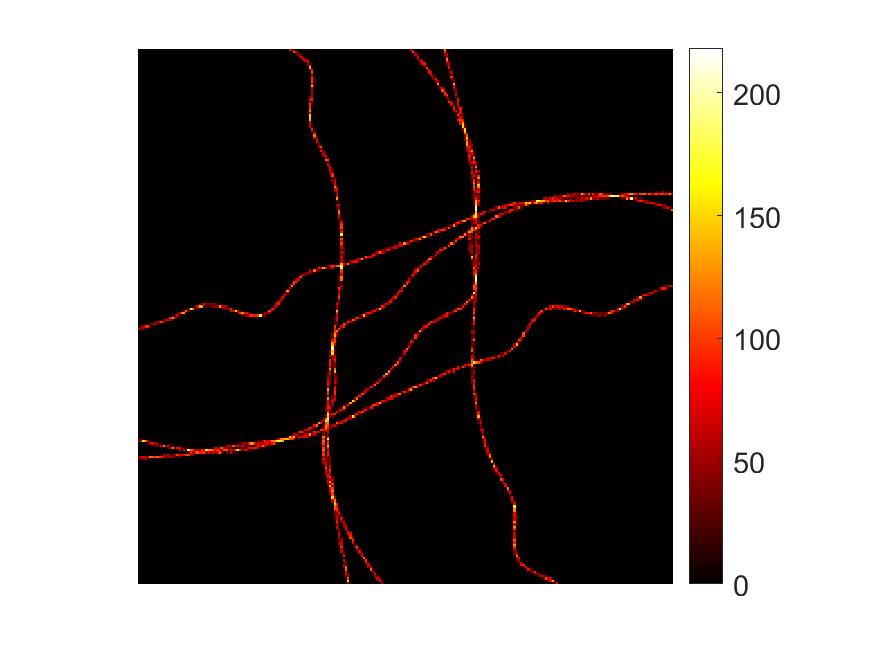}
     \caption{$\mathbf{x}_\text{GT}$}
     \label{fig:intensity_GT}
     \end{subfigure}

 \begin{subfigure}[b]{0.24\textwidth}
     \centering
     \includegraphics[width=\textwidth]{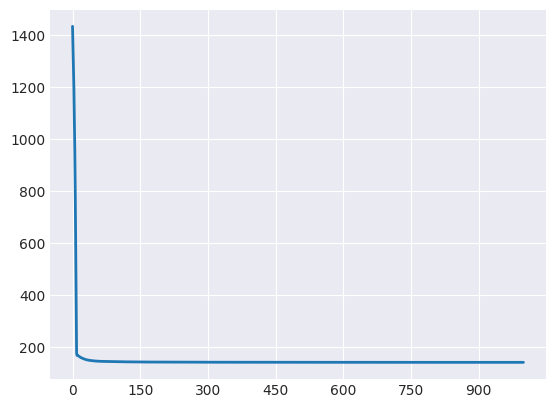}
     \caption{}
 \end{subfigure}
 \hfill
 \begin{subfigure}[b]{0.24\textwidth}
     \centering
     \includegraphics[width=\textwidth]{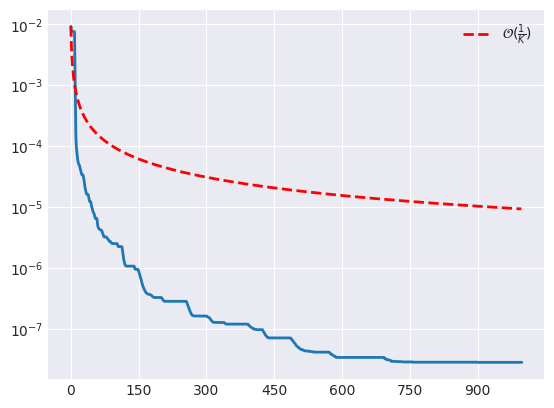}
     \caption{}
 \end{subfigure}
 \hfill
 \begin{subfigure}[b]{0.24\textwidth}
     \centering
     \includegraphics[height=2.8cm,trim={5cm 1cm  7cm 1cm},clip]{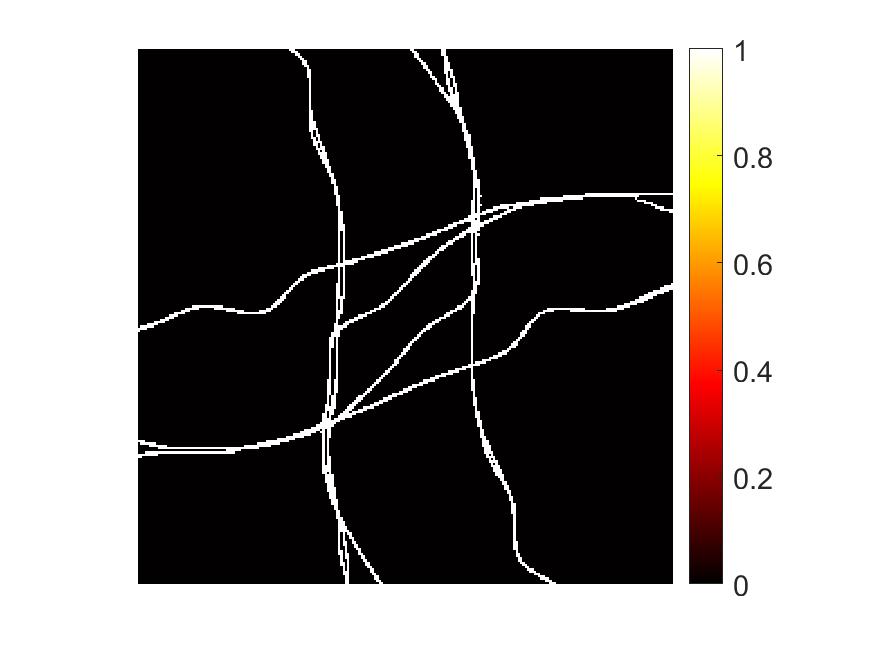}
     \caption{$\mathbf{\hat{\Omega}}$ (JI=0.67)}
     \label{fig:omega_hat}
 \end{subfigure}
 \hfill
 \begin{subfigure}[b]{0.24\textwidth}
         \centering
         \includegraphics[height=2.8cm,trim={5cm 1cm  3cm 1cm},clip]{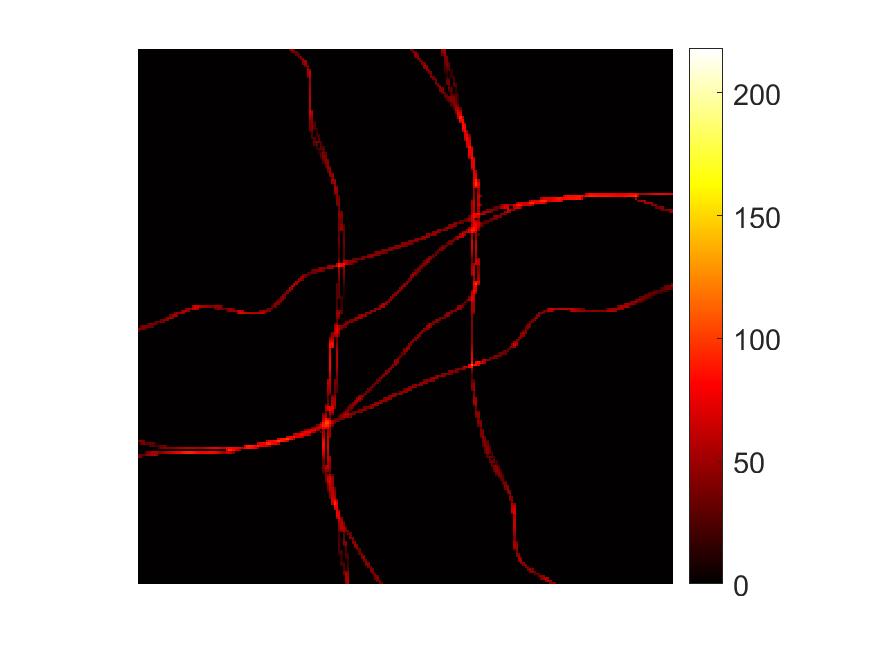}
     \caption{$\hat{\mathbf{x}}$ (PSNR=26.01)}
      \label{fig:intensity_estim}
     \end{subfigure}
     

\caption{(a)  Mean of the acquired temporal sequence, (b) First frame  (c) The ground truth support  (d) The ground truth intensity  image (e) Evolution of cost function $F_{\sigma}$ in \eqref{function PnP2} (f) the evolution of {$\min\limits_{i\leq k}\|\mathbf{r_x}^{i+1} - \mathbf{r_x}^{i} \|^2/ \|\mathbf{r_x}^{0}\|^2$}, in logarithmic scale, (g) Reconstructed support (h) Reconstructed intensity image.}
\label{difficult data PnP}
\end{figure}

We first apply PnP COL0RME to simulated data presented in Figure \ref{difficult data PnP}. 
The PSF used to generate the data has a FWHM equal to $176.6$ nm, the pixel size is equal to $25$ nm and the images have a size of $256 \times 256$ pixels.


Thanks to its training, we observe that the proposed approach is able to capture the filaments' geometry fairly well. We observe that in comparison to the ground truth support in Fig.~\ref{fig:supp_GT}, the reconstruction in Fig.~\ref{fig:omega_hat} is rather accurate. \rev{For the evaluation of the localization precision the Jaccard Index (JI) has been used. It is a quantity in $[0,1]$ computed as the ratio between correct detections (CD) and the total (correct, false negatives false positive) decetions, i.e. $\text{JI}:=CD/ (CD+FN+FP)$, up to a tolerance $\delta>0$, measured in nm (see, e.g., \cite{smlm}). For the reconstruction in Fig. ~\ref{fig:omega_hat}, the tolerance precision was chosen $\delta = 40$ nm.
}
Moreover, by solving \eqref{eq: intensity estimation}, intensities can also be estimated with high precision, see Fig.~\ref{fig:intensity_estim}. \rev{However, for the challenging dataset in Figure \ref{difficult data PnP}, the appearance of small artefacts (e.g. incorrect duplication of filaments) due to the training dataset we built are observed. They could be potentially removed by retraining the model with more heterogeneous data.}

\subsection{Real data}\label{PnP real data}

\begin{figure}[ht!]
     \centering
     \begin{subfigure}[b]{0.49\textwidth}
         \centering
         \includegraphics[height=4cm,trim={5cm 1cm  3cm 1cm},clip]{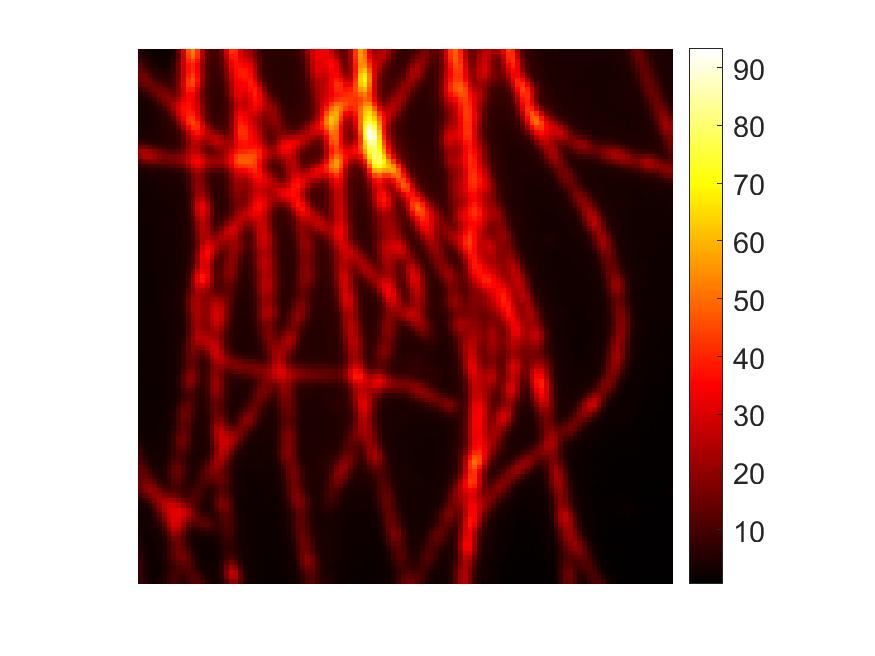}
         \caption{$\overline{\mathbf{y}}$}
     \end{subfigure}
     \hfill
      \begin{subfigure}[b]{0.49\textwidth}
         \centering
         \includegraphics[height=4cm,trim={5cm 1cm 3cm 1cm},clip]{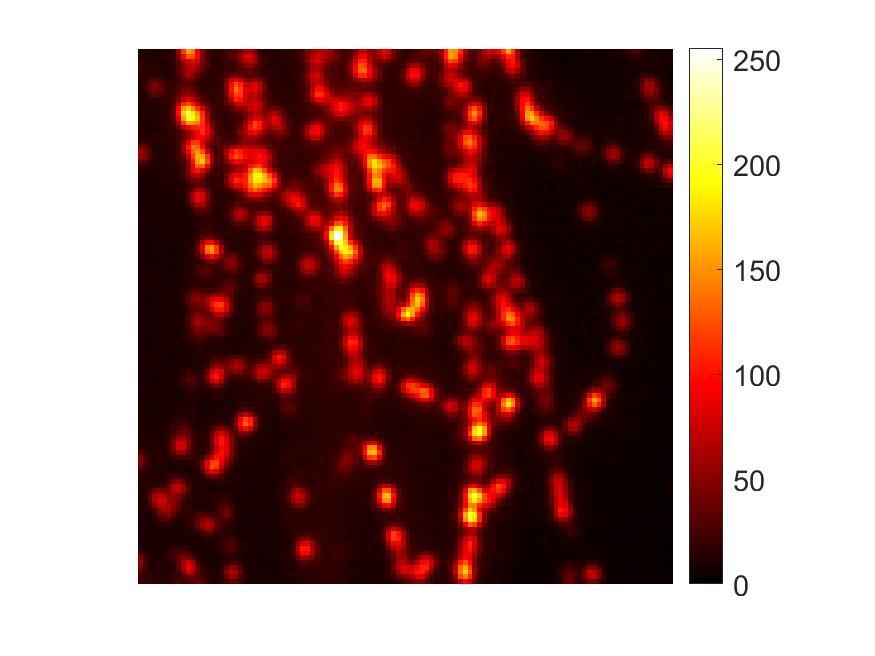}
         \caption{$\mathbf{y_1}$}
     \end{subfigure}

    \begin{subfigure}[b]{0.28\textwidth}
         \centering         \includegraphics[height=4cm,trim={5cm 1cm  7cm 1cm},clip]{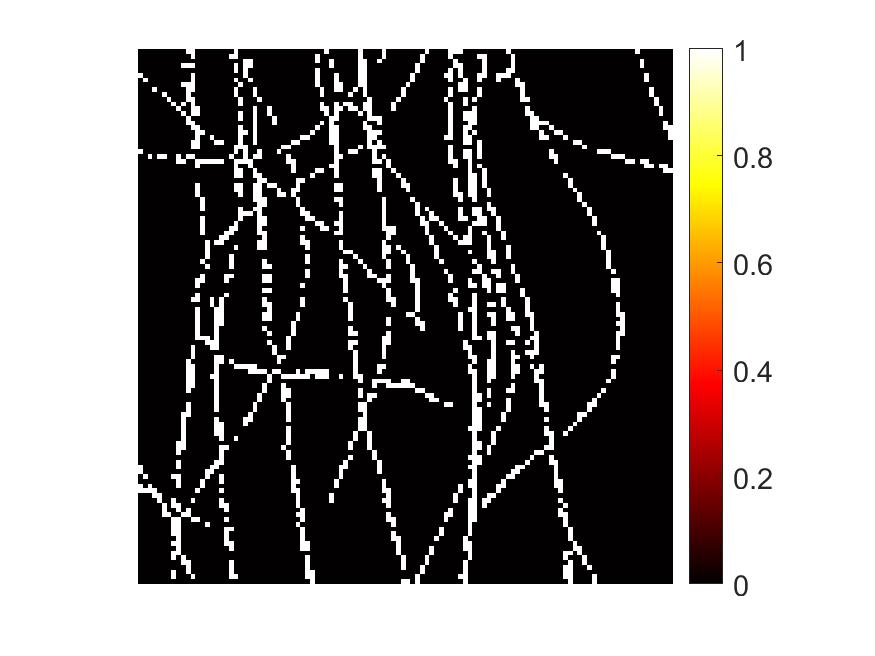}
         \caption{$\mathbf{\hat{\Omega}}$ by \cite{stergiopoulou_BioIm}}
     \end{subfigure}
    \hfill 
     \begin{subfigure}[b]{0.28\textwidth}
         \centering         \includegraphics[height=4cm,trim={5cm 1cm  7cm 1cm},clip]{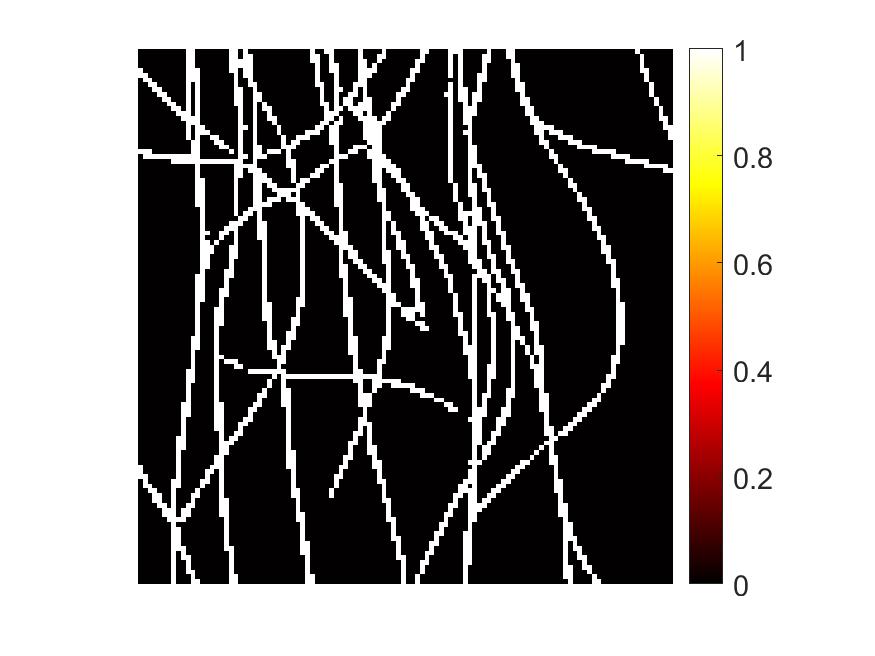}
         \caption{$\mathbf{\hat{\Omega}}$ by PnP Algo. \ref{Algorithm:AMA_support_auto}}
     \end{subfigure}
     \hfill
     \begin{subfigure}[b]{0.34\textwidth}
         \centering
         \includegraphics[height=4cm,trim={5cm 1cm 3cm 1cm},clip]{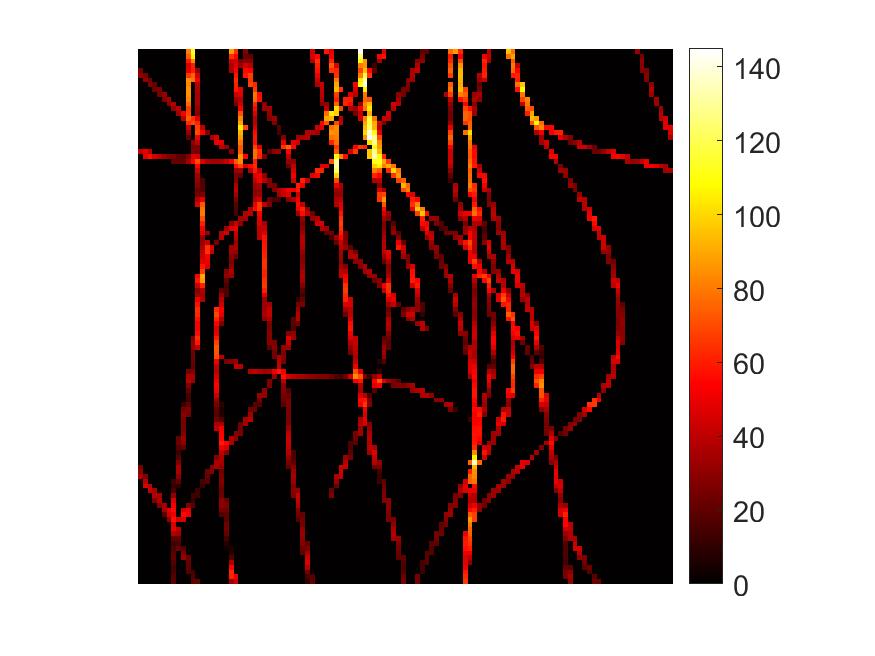}
         \caption{$\hat{\mathbf{x}}$}
    \end{subfigure}
    \caption{ HD-SMLM data: (first row) The temporal mean and the first frame of the acquired temporal sequence (second row) Support ($\ell_0$-based \cite{stergiopoulou_BioIm} VS.~PnP) and intensity reconstruction, $\sigma=10/255$.}
    \label{smlm fig}
    \end{figure}

We then applied the proposed approach to high-density SMLM acquisitions using a publicly available dataset created for the 2013 SMLM challenge \footnote{\href{https://srm.epfl.ch/Challenge/Challenge2013}{https://srm.epfl.ch/Challenge/Challenge2013}}, see Figure \ref{smlm fig}. Although in SMLM the molecules do not have a blinking behaviour, but rather an on-to-off transition, we can consider as blinking the temporal behaviour of one pixel in high-density videos due to the presence of many molecules per pixel. 
The dataset contains $T=500$ images, the PSF of the microscope used to acquire these data has a FWHM of $351.8$ nm and the pixel size is equal to $100$ nm. The support $\mathbf{\hat{\Omega}}$ computed by the model-based COL0RME approach in \cite{Stergiopoulou_ISBI21,stergiopoulou_BioIm} based on the use of a relaxation of the $\ell_0$ pseudo-norm \rev{is compared to} the one PnP-COL0RME variant of Algorithm \ref{Algorithm:AMA_support_auto}. Since no ground truth is available for these data, no quantitative assessment can be computed, however better continuation properties than COL0RME \cite{stergiopoulou_BioIm} are observed.

\section{Conclusions}

We presented a PnP model for support localisation for the  deconvolution of imaging data in fluorescence microscopy. PnP approaches rely on the use of off-the-shelf denoisers to model implicit prior regularisation functionals. They can be effectively used to replace proximal steps in proximal gradient algorithms. Following \cite{Hurault2022proximal}, we choose a denoiser with a particular structure to benefit from convergence guarantees. 
Our results show that the geometry of specific structures (thin filaments) can be captured by suitable training. 
Future work should take into account the presence of a downsampling operator in the image formation model and a more accurate modelling making use of also cross terms in the covariance data.

 {\small \section{Acknowledgements}
VS and LBF are supported by the 3IA
Côte d’Azur Investments (with reference number ANR-19-P3IA-0002). LC acknowledges the support received by the ANR project TASKABILE (ANR-22-CE48-0010) and the GdR ISIS project SPLIN.  VS, LC and LBF acknowledge the support received by the ANR project  MICROBLIND (ANR-21-CE48-0008). All authors acknowledge the support received by the H2020 RISE projects NoMADS (GA. 777826).
}


%
%
%

\vspace{-0.2cm}

\bibliographystyle{splncs04}
\bibliography{biblio}
%




\end{document}